\documentclass[aps,prl,twocolumn,superscriptaddress,groupedaddress]{revtex4}  
\usepackage{graphicx}  
\usepackage{dcolumn}   
\usepackage{bm}        
\usepackage{amssymb}   
\usepackage{amsmath}
\begin{document}

\widetext
\leftline{Primary authors: e.a.kozyrev@inp.nsk.su}
\leftline{To be submitted to PLB.}


\title{Study of the process $e^+ e^- \to K^0_{S}K^0_{L}$ 
in the center-of-mass energy range 1004--1060 MeV
with the CMD-3 detector at the VEPP-2000 $e^+ e^-$ collider.}   
\affiliation{Budker Institute of Nuclear Physics, SB RAS, 
Novosibirsk, 630090, Russia}
\affiliation{Novosibirsk State University, Novosibirsk, 630090, Russia}
\affiliation{Novosibirsk State Technical University, 
Novosibirsk, 630092, Russia}
\affiliation{Department of Physics and Astronomy, P.O. Box 3055 Victoria, B.C.,
CANADA, V8W 3P6}
\author{%
E.A.Kozyrev$^{1,2}$,
\quad E.P.Solodov$^{1,2}$,
\quad A.N.Amirkhanov$^{1,2}$,
\quad A.V.Anisenkov$^{1,2}$,
\quad V.M.Aulchenko$^{1,2}$,
\quad V.S.Banzarov$^{1}$,
\quad N.S.Bashtovoy$^{1}$,
\quad D.E.Berkaev$^{1,2}$,
\quad A.E.Bondar$^{1,2}$,
\quad A.V.Bragin$^{1}$,
\quad S.I.Eidelman$^{1,2}$,
\quad D.A.Epifanov$^{1,2}$,
\quad L.B.Epshteyn$^{1,2,3}$,
\quad A.L.Erofeev$^{1,2}$,
\quad G.V.Fedotovich$^{1,2}$,
\quad S.E.Gayazov$^{1,2}$,
\quad A.A.Grebenuk$^{1,2}$,
\quad S.S.Gribanov$^{1,2}$,
\quad D.N.Grigoriev$^{1,2,3}$,
\quad F.V.Ignatov$^{1}$,
\quad V.L.Ivanov$^{1,2}$,
\quad S.V.Karpov$^{1}$,
\quad A.S.Kasaev$^{1}$,
\quad V.F.Kazanin$^{1,2}$,
\quad A.N.Kirpotin$^{1}$,
\quad A.A.Korobov$^{1,2}$,
\quad O.A.Kovalenko$^{1,2}$,
\quad A.N.Kozyrev$^{1,2}$,
\quad I.A. Koop $^{1}$,
\quad P.P.Krokovny$^{1,2}$,
\quad A.E.Kuzmenko$^{1,2}$,
\quad A.S.Kuzmin$^{1,2}$,
\quad I.B.Logashenko$^{1,2}$,
\quad P.A.Lukin$^{1,2}$,
\quad K.Yu.Mikhailov$^{1,2}$,
\quad V.S.Okhapkin$^{1}$,
\quad A.V.Otboev$^{1}$,
\quad Yu.N.Pestov$^{1}$,
\quad A.S.Popov$^{1,2}$,
\quad G.P.Razuvaev$^{1,2}$,
\quad A.A.Ruban$^{1}$,
\quad N.M.Ryskulov$^{1}$,
\quad A.E.Ryzhenenkov$^{1,2}$,
\quad A.I.Senchenko$^{1}$,
\quad V.E.Shebalin$^{1,2}$,
\quad D.N.Shemyakin$^{1,2}$,
\quad B.A.Shwartz$^{1,2}$,
\quad D.B.Shwartz$^{1,2}$,
\quad A.L.Sibidanov$^{4}$,
\quad P.Yu.Shatunov$^{1}$,
\quad Yu.M.Shatunov$^{1}$,
\quad V.M.Titov$^{1}$,
\quad A.A.Talyshev$^{1,2}$,
\quad A.I.Vorobiov$^{1}$,
\quad Yu.V.Yudin$^{1,2}$
}

\date{\today}
\begin{abstract}
\hspace*{\parindent}
The $e^+ e^- \to K^0_{S}K^0_{L}$ cross section has been measured in 
the center-of-mass energy range 1004--1060 MeV at 25 energy points 
using $6.1 \times 10^5$ events with  $K^0_{S}\to \pi^+\pi^-$ decay. 
The analysis is based on 5.9 pb$^{-1}$ of an integrated luminosity
collected with the CMD-3 detector at the VEPP-2000 $e^+ e^-$ collider. 
To obtain $\phi(1020)$ meson parameters the measured cross section is 
approximated according to the Vector Meson Dominance model as a sum of the 
$\rho, \omega, \phi$-like amplitudes and their excitations.  This is the 
most precise measurement of the
$e^+ e^- \to K^0_{S}K^0_{L}$ cross section with a 1.8\% systematic 
uncertainty. 
\end{abstract}
\maketitle

\setcounter{footnote}{0}
\section{Introduction}
\label{Introd}
Investigation of $e^+ e^-$ annihilation into hadrons at low energy provides
unique information about interactions of light quarks.
High-precision studies of various hadronic cross sections are of great
interest in connection with the problem of the muon anomalous magnetic 
moment~\cite{g2} and constitute  the main goal of 
experiments with the CMD-3 and SND detectors at the upgraded
VEPP-2000 collider~\cite{snd,cmd}.

 In particular, $e^+ e^- \to K^0_{S}K^0_{L}$ is one of the processes with a 
rather large cross section in the center-of-mass energy range from
1 to 2 GeV. A precise measurement of this cross section,  
dominated by the contribution of the $\phi(1020)$ and 
$\phi(1680)$ resonances, is required to improve our knowledge of the
hadronic contributions to (g-2)$_\mu$ and $\alpha$(M$^2_Z$). 
Additional motivation for high-precision 
measurements of the $e^+ e^- \to K^0_{S}K^0_{L}$ and  $e^+ e^- \to K^+K^-$ 
cross sections around the $\phi$ meson peak comes from a significant 
deviation of the ratio of the coupling constants 
$\frac{g_{\phi \to K^{+}K^{-}}}{g_{\phi \to K_{S}K_{L}}}$ from 
theoretical predictions~\cite{Bramon}.
  
The most precise previous studies of the process have been performed at the 
CMD-2~\cite{cmdn}, SND~\cite{sndn} and BaBar~\cite{babarn} detectors. 
In this paper we present results of the new measurement of the 
$e^+ e^- \to K^0_{S}K^0_{L}$ cross section based on a high-statistics data
sample collected at 25 energy points in the center-of-mass energy (c.m.)
$E_{\rm c.m.}$ range 1004--1060 MeV with the CMD-3 detector.

\section{CMD-3 detector and data set}
\label{CMD3}

The Cryogenic Magnetic Detector (CMD-3) described elsewhere~\cite{cmd3} is installed in one of the two interaction regions of the VEPP-2000  $e^+ e^-$ collider~\cite{vepp2000000}.
The detector tracking system consists of the cylindrical drift chamber (DC) 
and double-layer cylindrical multiwire proportional Z-chamber, both installed inside a thin (0.085 $X_{0}$) superconducting solenoid with 1.3 T magnetic field. DC contains 
1218 hexagonal cells and provides a measurement of charged particle momentum 
and  of the polar ($\theta$) and azimuthal ($\phi$) angles.
An amplitude information from the DC wires is used to measure the 
ionization losses $dE/dx$ of charged particles with 
$ {\sigma}_{dE/dx}\approx$ 11-14\% accuracy for minimum ionization particles 
(m.i.p.).
A barrel electromagnetic calorimeter placed outside the solenoid consists of two subsystems: an inner liquid xenon (LXe) calorimeter (5.4 $X_{0}$ thick) surrounded by a scintillation CsI crystal calorimeter (8.1 $X_{0}$ thick)~\cite{csi}. 
BGO crystals with 13.4 $X_{0}$ are used as an endcap calorimeter.
The detector has two triggers: neutral and charged. A signal for neutral one is generated by the information from calorimeters, while the charged trigger 
comes from the tracking system.
The return yoke of the detector is surrounded by scintillation counters which 
veto cosmic events. 

To obtain a detection efficiency, Monte Carlo (MC) simulation of the detector 
based on the GEANT4~\cite{GEANT4} package has been developed. Simulated 
events are subject to the same reconstruction and selection procedures as 
the data. MC simulation includes photon jet radiation by initial electrons 
calculated according to Refs.~\cite{PJGen_sibid,rmc}. 
Background was estimated using a multihadronic Monte Carlo 
generator~\cite{gener}
based on experimental data for all measured processes in the
energy range up to 2 GeV.

The analysis uses 5.9 pb$^{-1}$ of an integrated luminosity
collected in two scans of the $\phi(1020)$ resonance region at 25 
energy points in the   $E_{\rm c.m.}$=1004--1060 MeV range.
The beam energy $E_{\rm beam}$ has been monitored by using the 
Back-Scattering-Laser-Light system~\cite{compton,compton1} which determines 
$E_{\rm c.m.}$ at each energy point with about 0.06 MeV accuracy.  
\section{Event selection}
\begin{figure}[h]
	\begin{center}
		\includegraphics[width=65mm]{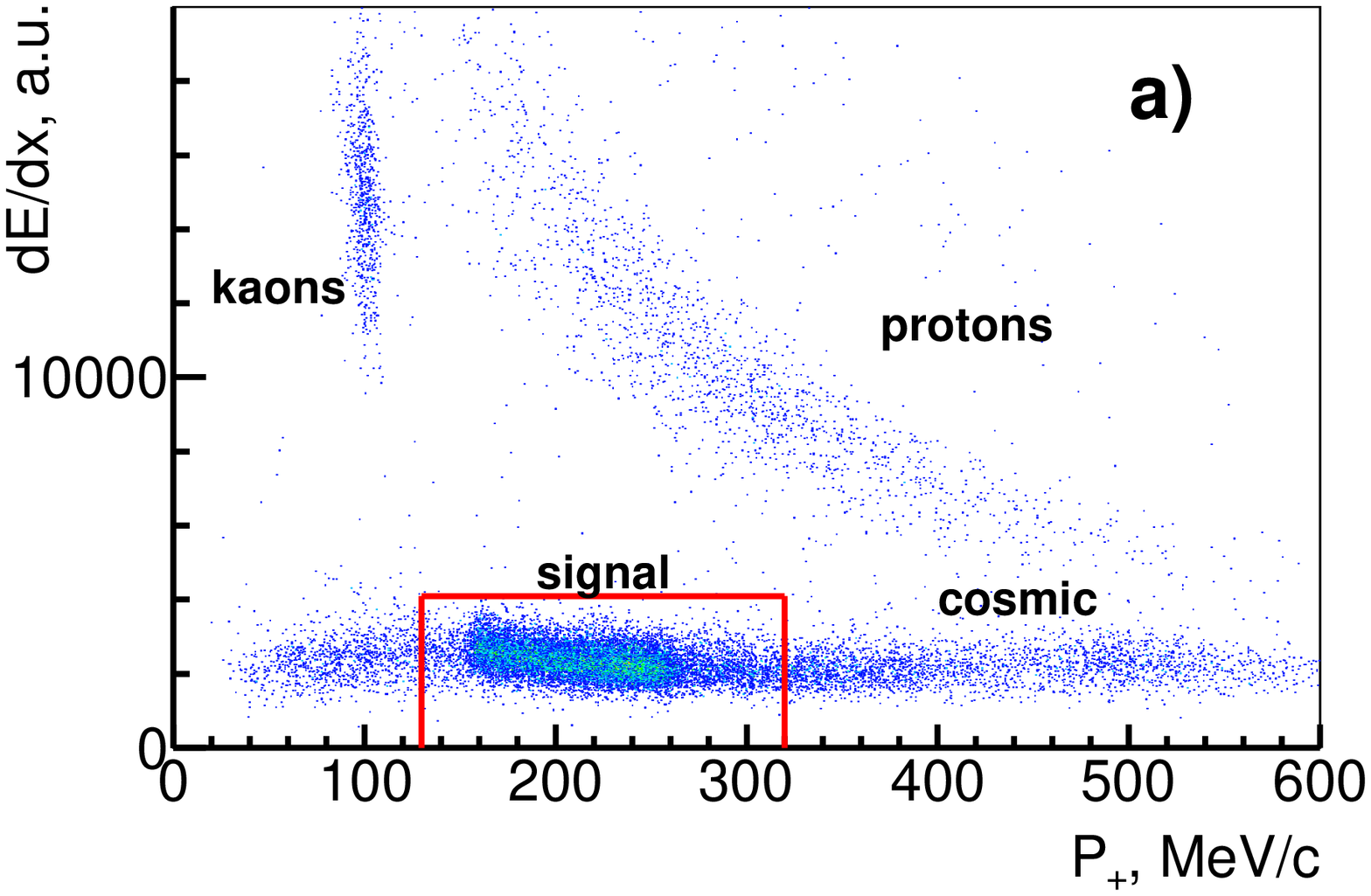}
                \includegraphics[width=65mm]{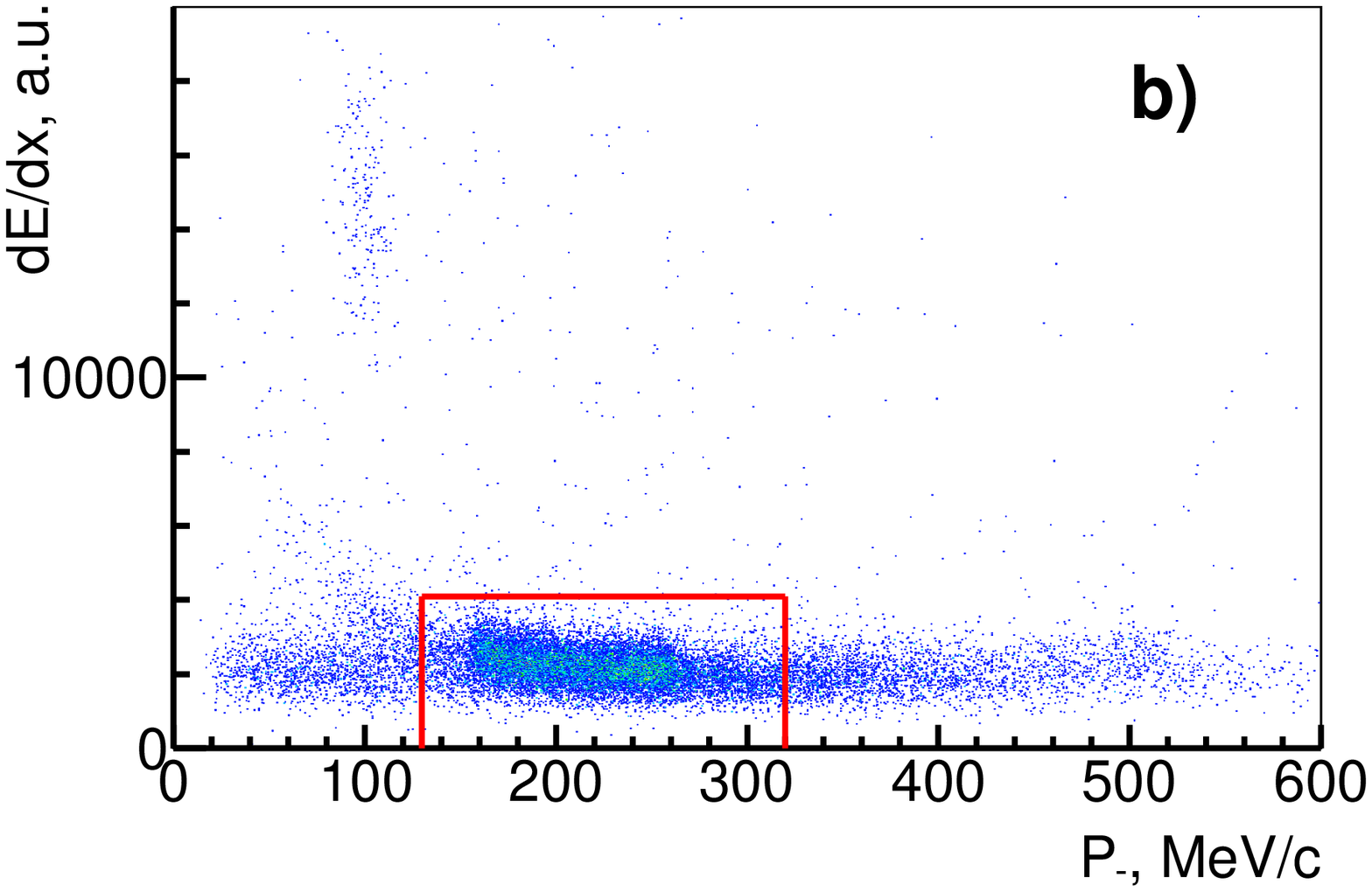}
		\caption{Ionization losses vs momentum  for positive (a) 
and negative (b) tracks for data at $E_{\rm beam} = 505$ MeV. The lines show 
selections of pions from the $K^0_{S}$ decay. 
		\label{dedxp}}
	\end{center}
\end{figure}

\begin{figure}[h]
	\begin{center}
		\includegraphics[width=65mm]{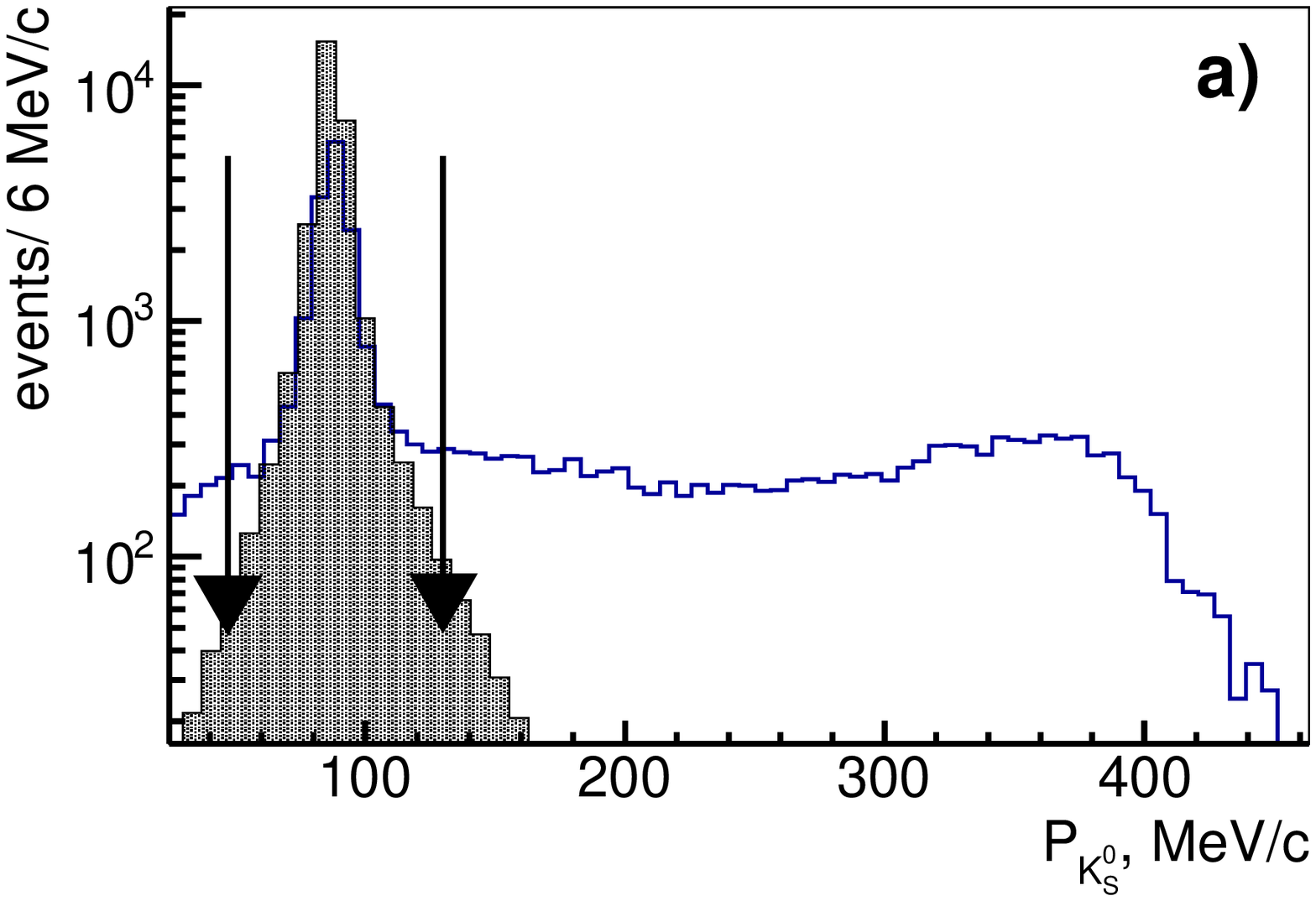}
                \includegraphics[width=65mm]{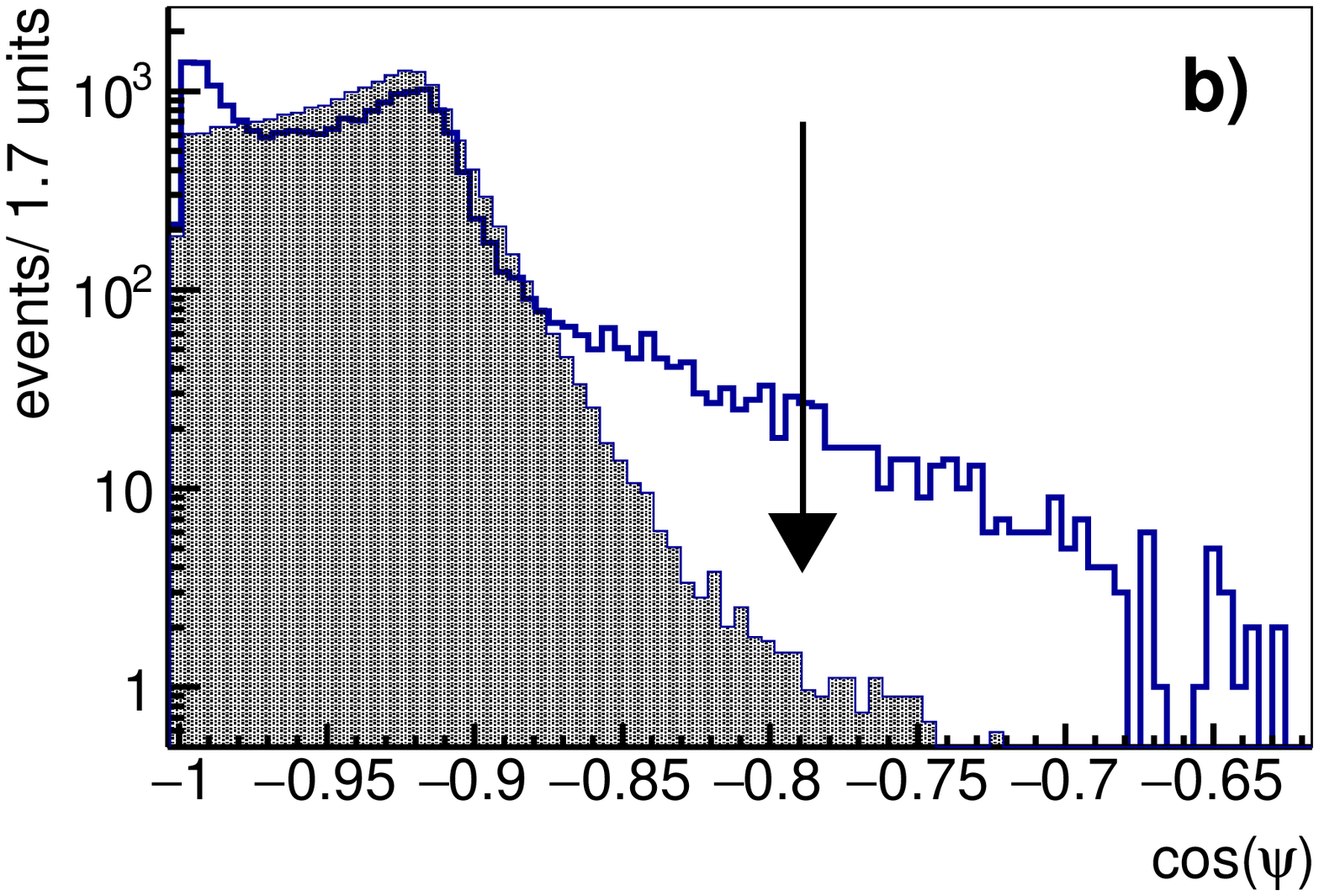}
		\caption{Total momentum  $P_{K_{S}^0}$ (a) and cosine of the 
angle $\psi$ between the two charged pions (b)  for the $K^0_{S}$ candidates 
after preliminary selection for data (open histogram) and MC simulation 
(shaded histogram) at $E_{\rm beam} = 505$ MeV. The arrows show additional selection requirements.
		\label{Ptot_508}}
	\end{center}
\end{figure}

Signal identification is based on detection of two pions from the 
$K^0_{S}\to \pi^+\pi^-$ decay. 
For each pair of oppositely charged tracks  a constrained fit to a common 
vertex is performed to determine  track parameters. 
Assuming tracks to be pions, the pair with the best $\chi^2$ from the 
vertex fit and with the invariant mass in the range 420--580 MeV/c$^2$ is 
selected as a $K_S^0$ candidate. The following requirements are applied to 
events with a found $K_S^0$ candidate:
\begin{trivlist}
\item $\bullet$ The longitudinal distance and the transverse coordinate of the vertex 
should have $|Z_{K_S^0}| < 10$ cm and $|\rho_{K_S^0}| < 6$ cm, 
respectively;

\item $\bullet$ Pions from $K_S^0$ decay are required to have polar angles 
1 $< \theta_{\pi^{+},\pi^{-}} < \pi$ - 1 radians;

\item $\bullet$ Each track has momentum 130 MeV/$c < P_{\pi^{\pm}} <$ 320 
MeV/$c$ corresponding to the kinematically allowed region for pions from the $K_S^0$ decay and its ionization losses in DC are within three standard deviations
from the average value, expected for pions. The last requirement rejects charged kaons and background protons, as shown in
Fig.~\ref{dedxp} for positive (a) and negative (b) tracks, respectively, at $E_{\rm beam} = 505$ MeV;

\item $\bullet$ The  momentum of the $K^0_{S}$ candidate, $P_{K_{S}^0} = |\vec{P}_{\pi^{+}} + \vec{P}_{\pi^{-}}|$, is required to be not larger than 
five standard deviations from the nominal momentum $P_{K_{S}^0} = \sqrt{E_{\rm c.m.}^2/4 - m_{K_{S}^0}^2}$ at each energy, as shown 
by the arrows in Fig.~\ref{Ptot_508}(a); 

\item $\bullet$ The cosine of the angle $\psi$ between the tracks should 
be smaller than the cosine of the minimal angle between two pions 
originating from the two-body decay of the $K_S^0$ meson, shifted by five 
standard deviations, as shown by the arrow in Fig.~\ref{Ptot_508}(b).
\end{trivlist}
 
\begin{figure}[h]
	\begin{center}
		\includegraphics[width=65mm]{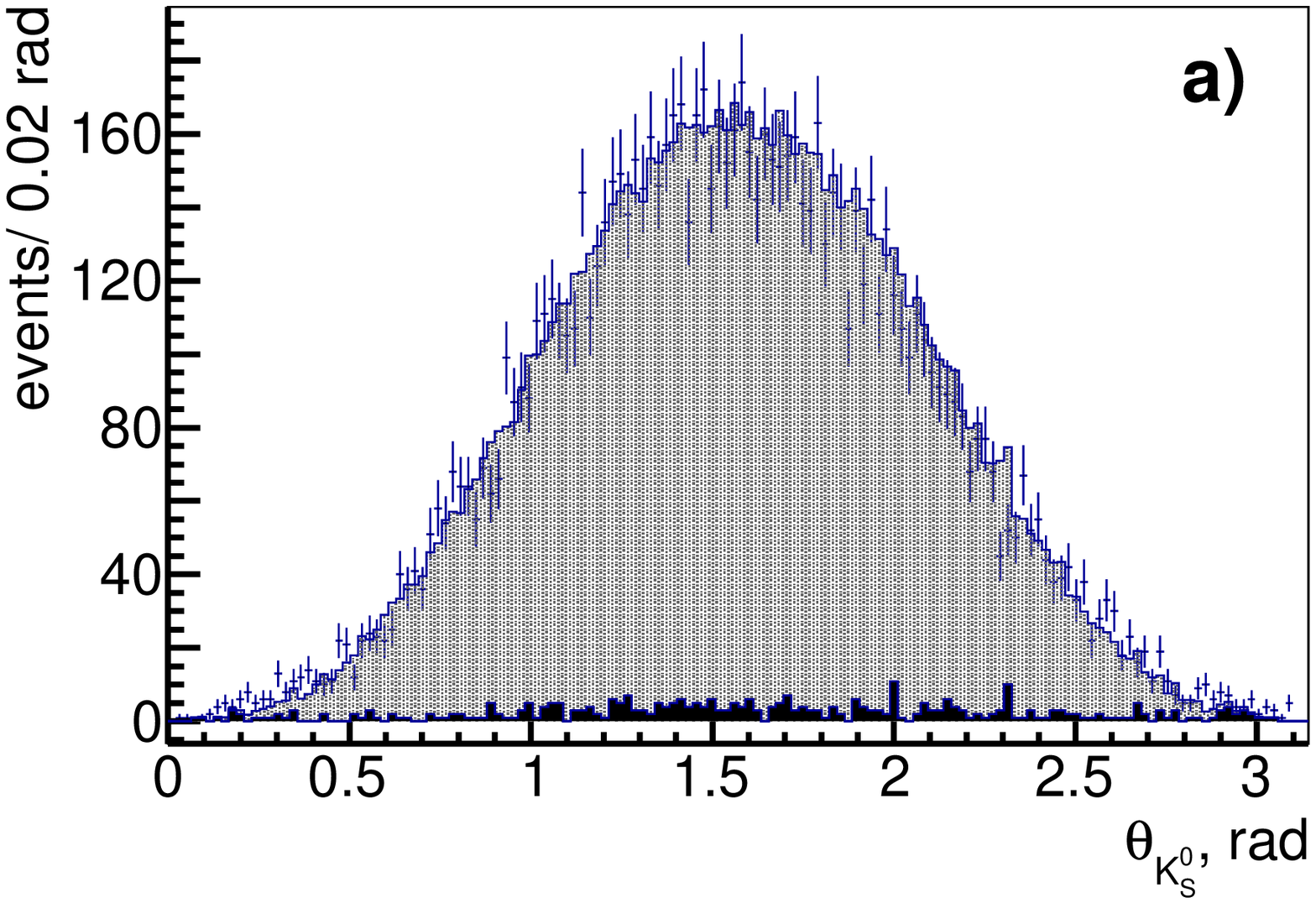}
                \includegraphics[width=65mm]{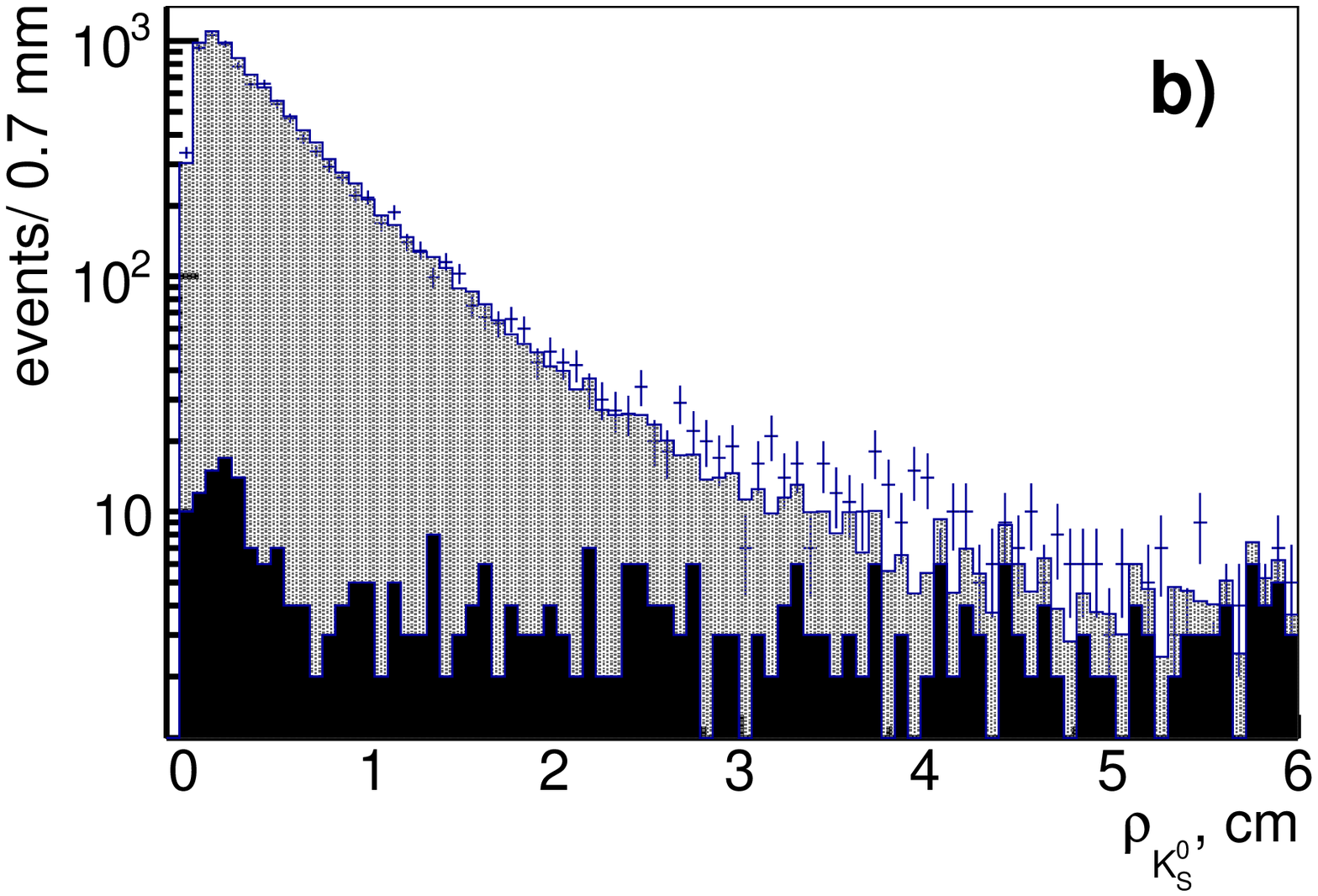}
		\caption{Reconstructed polar angle of the $K_S^0$ meson (a) 
and the transverse distance of the $K_S^0$ decay vertex from the beam (b) at 
$E_{\rm beam} = 505$ MeV for data (points) and  signal simulation (shaded 
histogram). The dark shaded histograms represent the estimated contribution 
from  the  background processes.
		\label{Minv_508}}
	\end{center}
\end{figure}

The reconstructed polar angle of the $K_S^0$ meson and the transverse distance
of the $K_S^0$ decay vertex from the $e^+ e^-$ interaction point are shown 
in Fig.~\ref{Minv_508} after above selections for data (points) and 
MC-simulation (shaded histogram).
The dark shaded histograms show a sum of the background contributions from 
the MC-simulated hadronic processes (predominantly $e^+e^-\to\pi^+\pi^-2\pi^{0}$) and a contribution  from cosmic muons estimated using  
events from the  $|Z_{K_S^0}|$ sideband ($10 < |Z_{K_S^0}| < 15$ cm). 
%
%

\begin{figure}[h]
	\begin{center}
	     \includegraphics[width=65mm]{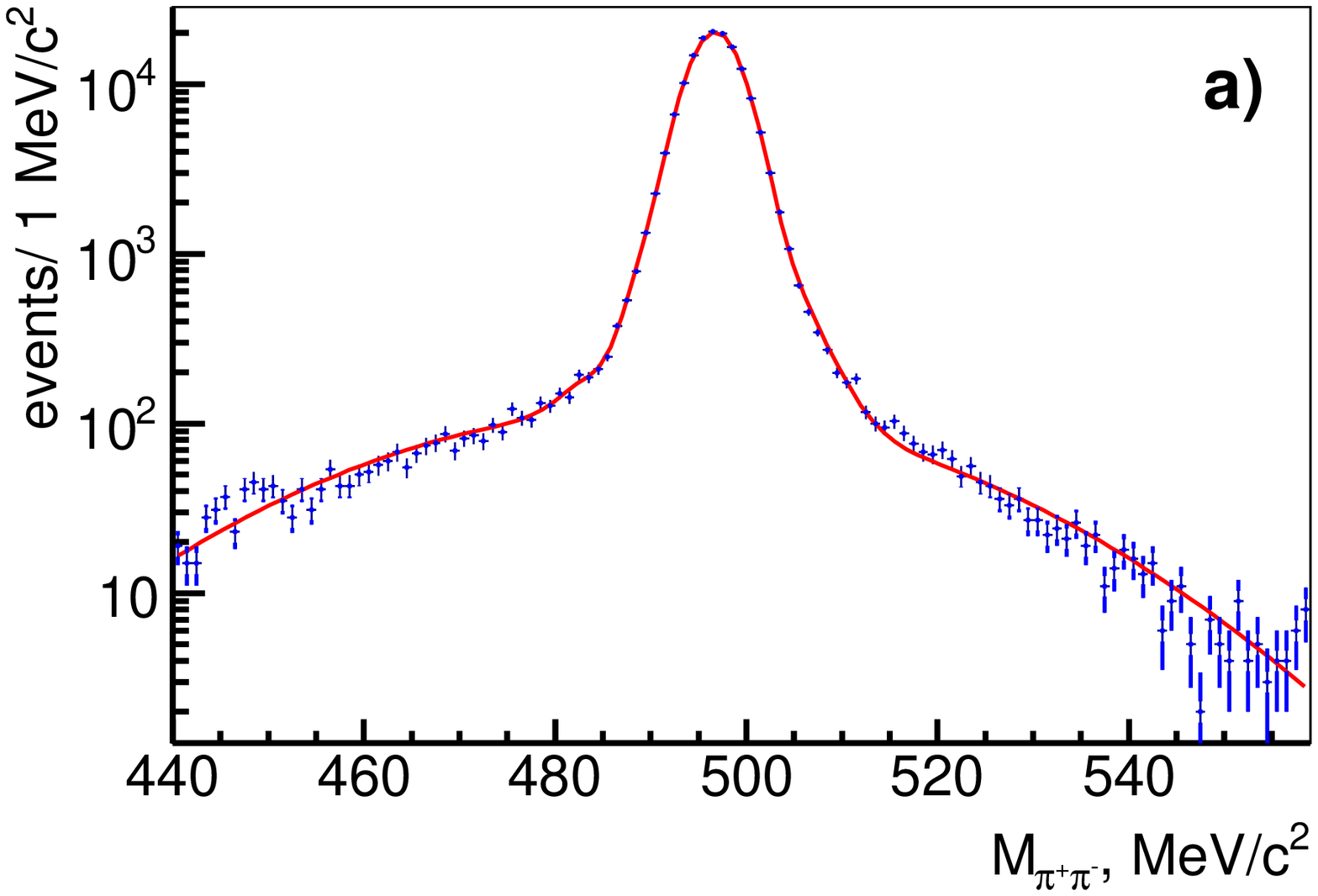}
		\includegraphics[width=65mm]{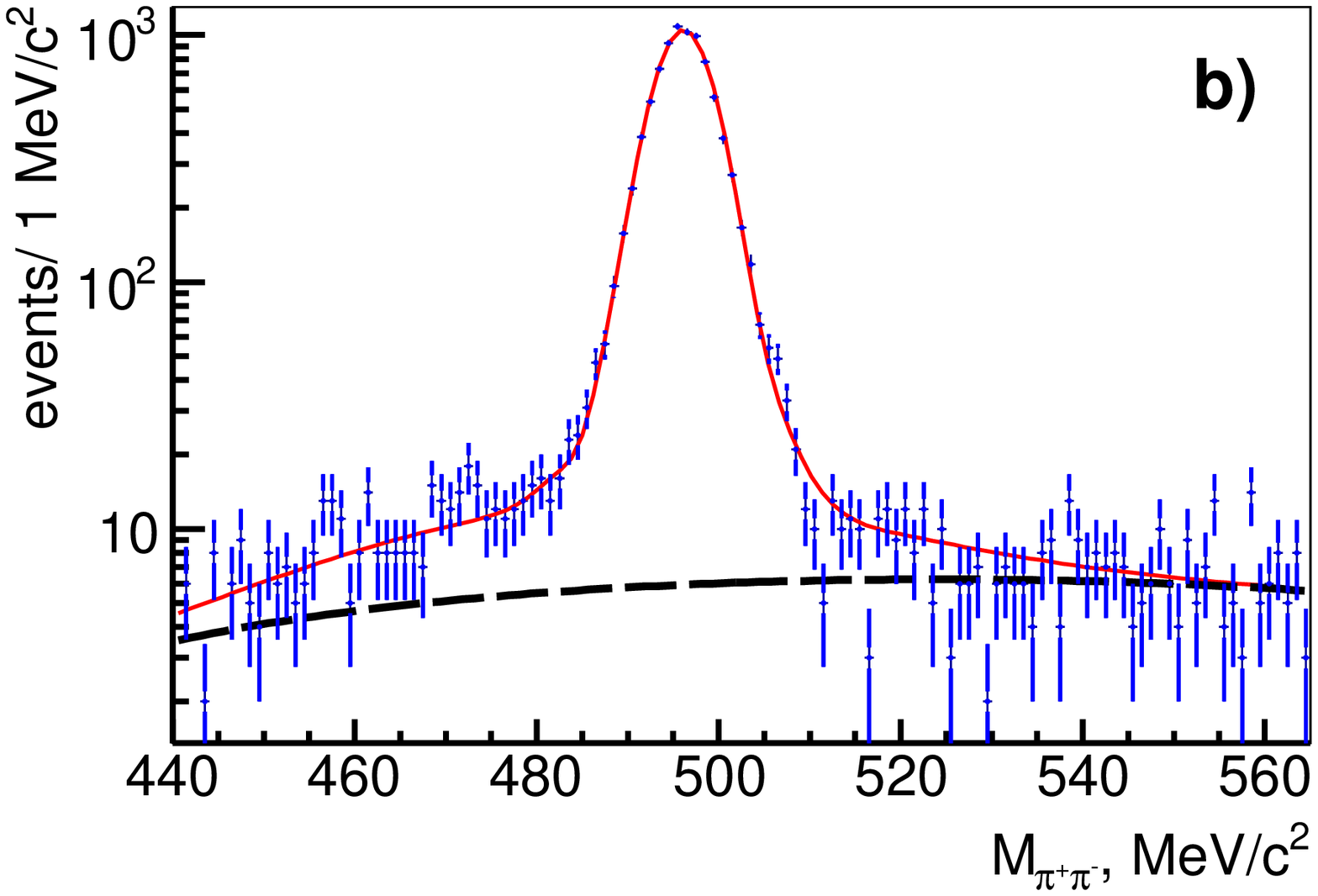}
		\caption{Approximation of the invariant mass of two pions at $E_{\rm beam} = 505$ MeV  for simulation (a) and data (b). The solid line 
corresponds to the signal, the long-dotted line to the background.
		\label{2pimass}}
	\end{center}
\end{figure}

We determine the number of signal events for data and simulation from a 
binned maximum likelihood fit of two-pion invariant mass shown in 
Fig.~\ref{2pimass}. The signal shape is described by a sum 
of four Gaussian functions with parameters fixed from the simulation 
and with additional Gaussian smearing to account for the difference 
in data-MC detector responses. 
The background in data, described by a second-order polynomial function, constitutes about 30\% outside the $\phi$ meson peak and 0.5\% under it.
By toy MC experiments with fixed signal and background profiles as well as 
by varying the background shape and approximation range used we estimate an uncertainty on the number of extracted signal events as less than 1.1\%.
The number of obtained signal events, $N_{\rm exp}$, for each energy  is listed 
in Table~\ref{endtable}. 

\section{Cross section of $e^+ e^- \to K^0_{S}K^0_{L}$ }
The Born cross section of the process $e^+ e^- \to K^0_{S}K^0_{L}$ is 
calculated at each energy from the expression: 
\begin{eqnarray}  
    \sigma^{\rm Born} &=& \frac{N_{\rm exp}}{\epsilon_{\rm reg}\epsilon_{\rm trig} L (1 + \delta^{\rm rad.})}(1+\delta^{\rm en.spr.}),  
\end{eqnarray}
where   
$\epsilon_{\rm reg}$ is a detection efficiency, 
$\epsilon_{\rm trig}$ is a trigger efficiency,
$L$ is an integrated luminosity, 
$1+\delta^{\rm rad.}$ is a radiative correction, and  
$1+\delta^{\rm en.spr.}$ represents a correction due to the spread of the
collision energy.

\begin{figure}[h]
\begin{minipage}[]{0.46\textwidth}
  \includegraphics[width=0.98\textwidth]{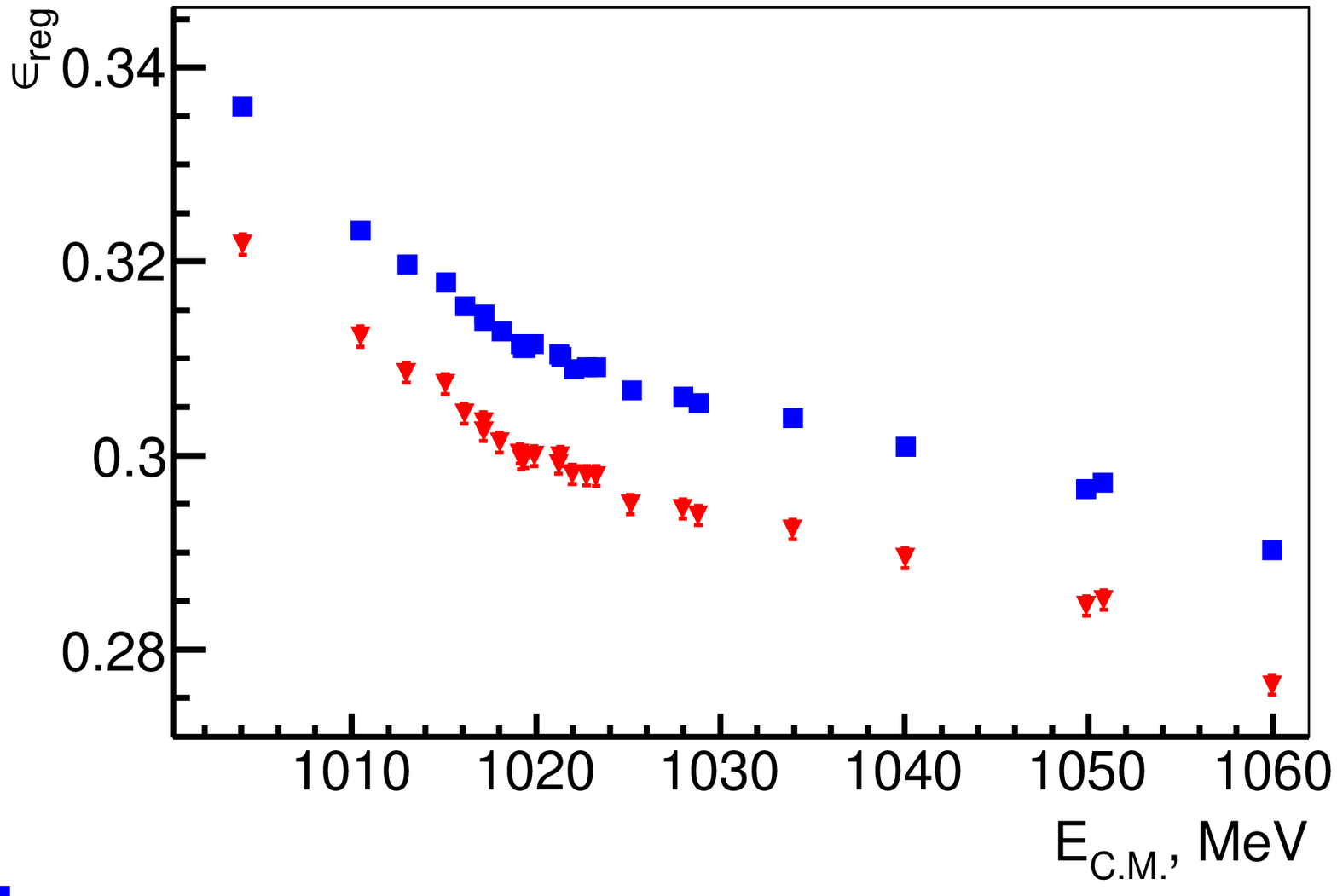}
  \caption{Detection efficiency of the $K^0_{S}K^0_{L}$ pairs vs energy 
from simulation (triangles). The geometrical efficiency is shown by squares (see text).
  }
  \label{efficiencypict}
\end{minipage}
\hfill
\begin{minipage}[]{0.46\textwidth}
		\includegraphics[width=0.98\textwidth]{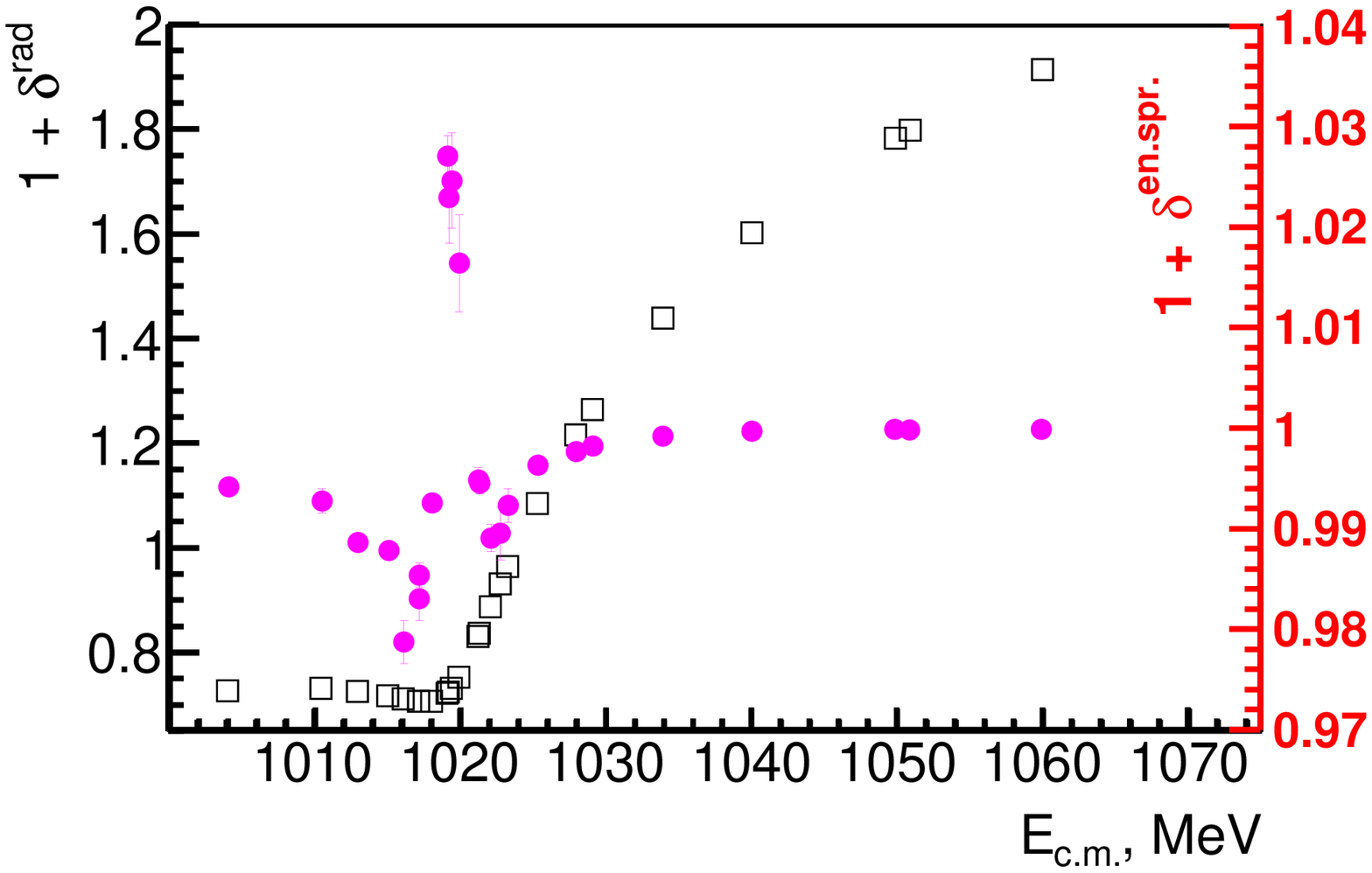}
		\caption{Radiative corrections $1+\delta^{\rm rad.}$ 
(squares, left scale) and corrections $1+\delta^{\rm en.spr.}$ for the spread of collision 
energy (points, right scale).
		\label{radpfig}}
\end{minipage}
\end{figure}

The detection efficiency $\epsilon_{\rm reg}$ is obtained by dividing 
the number of MC simulated events after reconstruction and selection 
described above by the total number of generated $K_{S}^0 K_{L}^0$ pairs 
taking into account the branching fraction 
$B_{K^0_{S}\to \pi^+\pi^-} = (69.20\pm0.05)\%$~\cite{PDG}. 
Figure~\ref{efficiencypict} shows the obtained detection efficiency 
(triangles) vs c.m. energy in comparison with the expected geometrical 
efficiency (squares).  The geometrical efficiency is calculated as the 
probability of pions to be in the polar angle range 
1 $< \theta_{\pi^{+},\pi^{-}} < \pi$ - 1 radians at the generator level. 
%

The trigger efficiency is studied using responses of two independent 
triggers, charged and neutral, for selected signal events, and is found to 
be close to unity, $\epsilon_{\rm trig} = 0.998 \pm 0.001$.

The integrated luminosity $L$ is determined using events of the processes 
$e^+e^- \to e^+e^-$ (Bhabha events) with about 1\%~\cite{lum} systematic accuracy.
\begin{figure}
\begin{center}
  \includegraphics[width=0.5\textwidth]{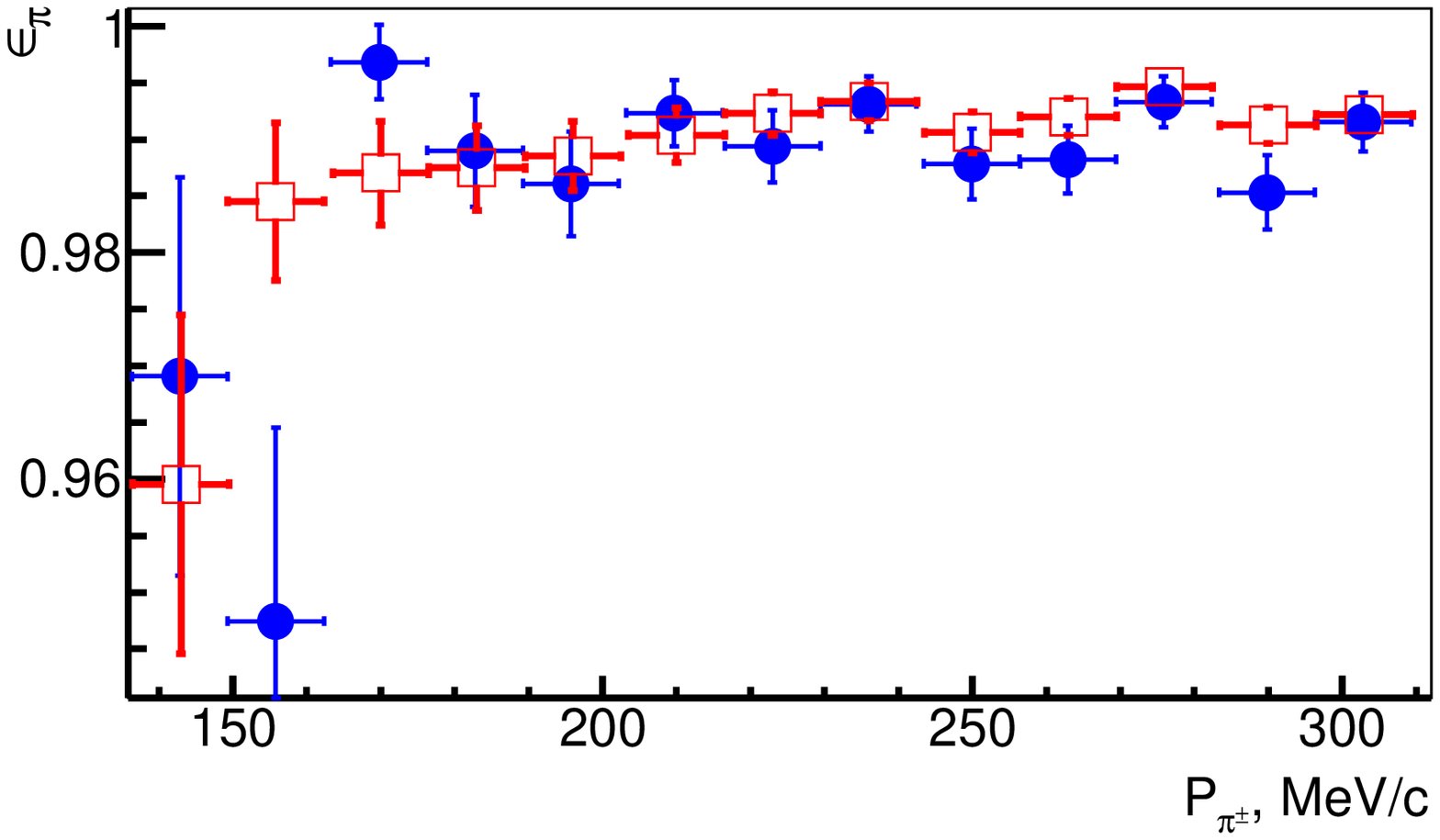}
  \caption{Pion detection efficiency in DC vs momentum for data 
(circles) and simulation (squares).}
  \label{piefficien}
\end{center}
\end{figure}

The initial-state radiative correction $1+\delta^{\rm rad.}$, shown 
by squares in Fig.~\ref{radpfig}, is calculated using the structure function 
method with an accuracy better than 0.1\%~\cite{radcorFadin}.

The spread of collision energy is about 350 keV, that is significant in 
comparison with the $\phi$ meson width, and we introduce the correction 
of the cross section, shown by points in Fig.~\ref{radpfig}, which 
has a maximum value of 1.028$\pm$0.004 at the peak of the $\phi$ resonance.

The resulting cross section is listed in Table~\ref{endtable} for each energy
and shown in Fig.~\ref{cross_phi}. 
The presented errors are statistical only and include fluctuations of signal 
and Bhabha events as well as the error  $\delta E_{\rm c.m.}$
due to the statistical uncertainty of the c.m. energy measurement.
The last part was calculated as $|\frac{\partial\sigma^{\rm Born}}{\partial E_{c.m.}}| \times \delta E_{\rm c.m.}$.

\begin{table}
\begin{center}
\caption{Summary of systematic uncertainties in the 
$e^+e^- \to K_{S}^0 K_{L}^0$ cross section measurement}
\label{endtable_syst}
\begin{tabular}[t]{||c|c||}
\hline
Source                        &   Uncertainty, \% \\
\hline
Signal extraction by fit      & 1.1         \\
Detection efficiency          & 1.0             \\
Radiative correction          & 0.1      \\
Energy spread correction      & 0.3 \\
Trigger efficiency            & 0.1 \\
Luminosity                    & 1.0   \\
\hline
Total                         & 1.8      \\
\hline
\end{tabular}
\end{center}
\end{table}
\section{Systematic uncertainties}
MC simulation may not exactly reproduce all detector responses, so an 
additional study was performed to obtain corrections for  data-MC difference 
in the detection efficiency. 

The data-MC difference in the charged pion detection by DC is studied using 
the process $e^+e^- \to \phi \to \pi^+\pi^-\pi^0$. Three-pion events 
can be fully reconstructed from  one detected charged track and 
two detected photons from the $\pi^0$ decay, and a probability to detect 
another charged track can be determined.
For the polar angle requirement 1 $< \theta_{\pi^{+},\pi^{-}} < \pi$ - 1 radians, 
the average detection inefficiency is about 1\% per track for high momentum, and decreases 
with pion momentum, as shown in Fig.~\ref{piefficien}. The rise of efficiency 
vs momentum is explained by the decreasing number of pions that decayed or 
interacted in DC. 
Good  data-MC agreement is observed for charged pion detection, so no 
efficiency correction is introduced and the uncertainty in the detection 
is estimated as 0.5\%.

DC calibration is checked using signals of the Bhabha events~\cite{lum} 
in the DC and Z-chamber, and  for pions from the 
$K_{S}^0$ decay the uncertainty due to the polar angle selection in the range 
of polar angles chosen is estimated as 0.4 \%. 

By variation of corresponding selection criteria we estimate the uncertainty 
due to the data-MC difference in the angular and momentum resolutions 
as 0.5\%, while other selection criteria contribute another 0.6\%. 

The total uncertainty of the detection efficiency is calculated as a 
quadratic sum of uncertainties from the different sources and is 
estimated to be 1.0\%.

The systematic uncertainties of the $e^+e^- \to K_{S}^0 K_{L}^0$ cross section 
discussed above are summarized in Table~\ref{endtable_syst} 
giving 1.8\% in total.

\section{Fitting of the $e^+ e^- \to K^0_{S}K^0_{L}$ cross section}
%
To obtain $\phi(1020)$ parameters
we approximate the energy dependence  of the cross section according to the 
vector meson dominance (VMD) model as a sum of the $\rho,~\omega,~\phi$-like 
amplitudes~\cite{Kuhn}:

\begin{equation}
\begin{aligned}
\sigma_{e^+ e^- \to K^0_{S}K^0_{L}}(s) = \frac{8 \pi \alpha}{3 s^{5/2}} p_{K^0}^{3} 
| \frac{g_{\rho \gamma} g_{\rho KK}}{D_{\rho}(s)} 
+ \frac{g_{\omega \gamma} g_{\omega KK}}{D_{\omega}(s)} \\
+ \frac{g_{\phi \gamma} g_{\phi KK}}{D_{\phi}(s)}
+ A_{\rho',\omega',\phi'}|^{2},
\label{ksklxs}
\end{aligned}
\end{equation}
where $s = E_{\rm c.m.}^2$, $p_{K^0}$ is a neutral kaon momentum, 
$D_{V}(s) = m_{V}^{2} -s - i \sqrt{s}\Gamma_{V}(s)$,
$m_{V},$ and $\Gamma_{V} $ are mass and width of the major intermediate 
resonances: $V = \rho(770),~\omega(782),$ $\phi(1020)$. 
The energy dependence of the decay width is expressed via a sum of partial 
widths multiplied by a factor of  phase space energy dependence 
$P_{V\to f}(s)$ of each decay mode as: 
\begin{eqnarray*}
\Gamma_{V}(s) = \Gamma_{V} \sum_{V\to f} B_{V\to f} \frac{P_{V\to f}(s)}{P_{V\to f}(m_V^2)}.
\end{eqnarray*}

The coupling constants of the intermediate vector meson $V$ with 
initial and final states can be presented as: 
\begin{eqnarray*}
|g_{V \gamma}| = \sqrt{\frac{3 m_{V}^{3} \Gamma_{Vee}}{4 \pi \alpha}};~|g_{V KK}| = \sqrt{\frac{6 \pi m_{V}^{2} \Gamma_{V} B_{VKK}}{p^{3}_{K^0}(m_{V})}},
\end{eqnarray*}
where $\Gamma_{Vee}$ and $B_{VKK}$ are electronic width and branching 
fraction of the $V$ meson decay to a pair of kaons.

In our approximation we use the world-average values of mass, total width 
and electronic width of the $\rho(770)$ and $\omega(782)$:  
$\Gamma_{\rho\to ee} = 7.04\pm0.06~\rm keV, 
\Gamma_{\omega\to ee} = 0.60\pm0.02~\rm keV$~\cite{PDG}.
The branching fractions of the $\rho(770)$ and $\omega(782)$ to a kaon pair 
are unknown, and 
we use the relation $g_{\omega K^0_{S}K^0_{L}} = - g_{\rho K^0_{S}K^0_{L}} = - g_{\phi K^0_{S}K^0_{L}}/\sqrt{2}$, based on the quark model with ``ideal" mixing and exact SU(3) symmetry 
of u-,d-,s-quarks~\cite{Kuhn}.

The amplitude  $A_{\rho',\omega',\phi'}$ denotes a contribution of excited 
$\rho(1450),~\omega(1420)$ and $\phi(1680)$ vector meson states in the 
$\phi(1020)$ mass region. Using BaBar~\cite{babarn} data above 1.06 GeV for the process $e^+e^- \to K^0_{S}K^0_{L}$  we found a relatively small contribution of these states in the studied energy range in comparison with nonresonant $\rho$ and $\omega$ contributions. 

\begin{figure*}[h]
\begin{center}
  \includegraphics[width=1.1\textwidth]{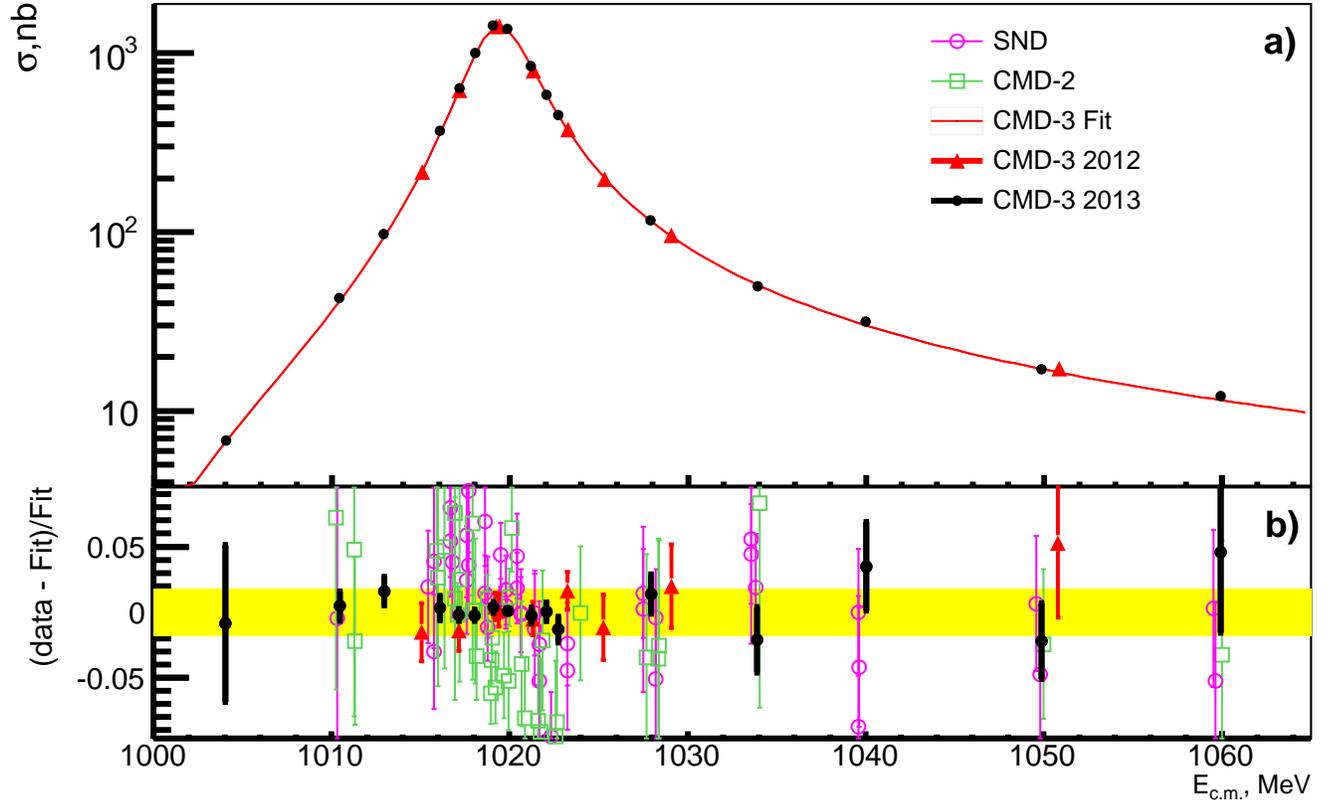}
  \caption{(a) Measured $e^+e^- \to K^0_{S}K^0_{L}$ cross section in 
comparison with previous experiments. The dots are experimental data, 
the curve is the fit described in the text. (b) Relative difference between the data and fit. Comparison with other experimental data is shown. Statistical uncertainties only are included for data. The width of the band shows the systematic uncertainties in our experiment.}
  \label{cross_phi}
\end{center}
\end{figure*}

We perform a fit to the $e^+e^- \to K_{S}^0 K_{L}^0$ cross section with floating 
$m_{\phi},~\Gamma_{\phi}$, and $\Gamma_{\phi \to ee}\times B_{\phi \to K^0_S K^0_L}$ 
(or alternatively $B_{\phi \to ee}\times B_{\phi \to K^0_S K^0_L}$) 
parameters: the fit yields $\chi^2/ndf = 20/22$ ($P(\chi^2)=58\%$).  The 
contributions of the $\rho$ and $\omega$ intermediate states
are non-negligible and we performed a fit where we introduce an additional 
floating parameter $g_{\rho,\omega}$, which 
is a multiplicative factor for both $g_{\omega K^0_{S}K^0_{L}}$ and 
$g_{\rho K^0_{S}K^0_{L}}$ coupling constants in Eq.~\ref{ksklxs}.
The fit yields $\chi^2/ndf = 15/21$  ($P(\chi^2)=82\%$)  
with $g_{\rho,\omega} = 0.80 \pm 0.09$.
This is the first quantitative estimate of the $\rho$ and $\omega$
amplitude contributions in the $\phi$ meson region.
The obtained parameters of the $\phi$ meson in comparison with the 
values of other measurements 
are presented in Table~\ref{endtable_phi_parameters} 
and the fit result is shown in Fig.~\ref{cross_phi}(a). Figure~\ref{cross_phi}(b) shows 
the relative difference between the 
obtained data and the fit curve. Only statistical uncertainties are shown.
The width of the band shows the systematic uncertainty in our measurement. A slope of the CMD-2 points~\cite{cmdn} can be explained by an about 80 keV difference between the used values of c.m. energy in the previous work and this experiment, that is within declared systematic uncertainties of the energy measurements.
 
\begin{figure*}[h]
\begin{center}
  \includegraphics[width=0.8\textwidth]{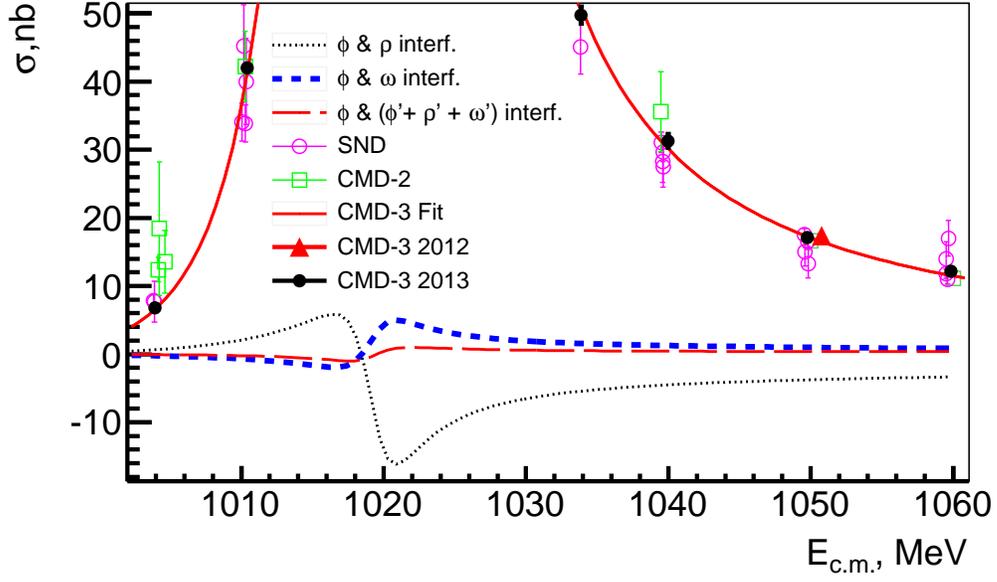}
  \caption{Contributions of lower- and higher-mass resonances to the fit of the $e^+e^- \to K^0_{S}K^0_{L}$ cross section in the studied energy range.}
  \label{cross_interf}
\end{center}
\end{figure*}

The  contributions of the $\rho$ and $\omega$ intermediate states 
are demonstrated in Fig.~\ref{cross_interf} by the dotted lines, while 
the long-dashed line shows a contribution from higher excitations.  The first 
uncertainties presented in Table~\ref{endtable_phi_parameters} are 
statistical, and the second are the systematic uncertainties.
Two effects were taken into account in the estimation of the latter: 
the accuracy of the measurement of the c.m.s. energy $E_{c.m.}$ of 60 keV and 
the systematic uncertainty of the cross section measurement of 1.8\% (Table~\ref{endtable_syst}). 
To study model dependence of the results, 
several additional fits are performed.
Other fits use Eq.~\ref{ksklxs} without the $A_{\phi',\rho',\omega'}$ amplitude 
and introduce an additional floating phase of the $\phi$ meson 
amplitude or the both $\rho$ and $\omega$ amplitudes. The variations in the $\phi$ meson parameters are used as an 
estimate of the model-dependent uncertainty presented as a third uncertainty 
in Table~\ref{endtable_phi_parameters}.
The obtained values agree with results of other 
measurements and some are more precise.

\begin{figure*}[h]
\begin{center}
  \includegraphics[width=1\textwidth]{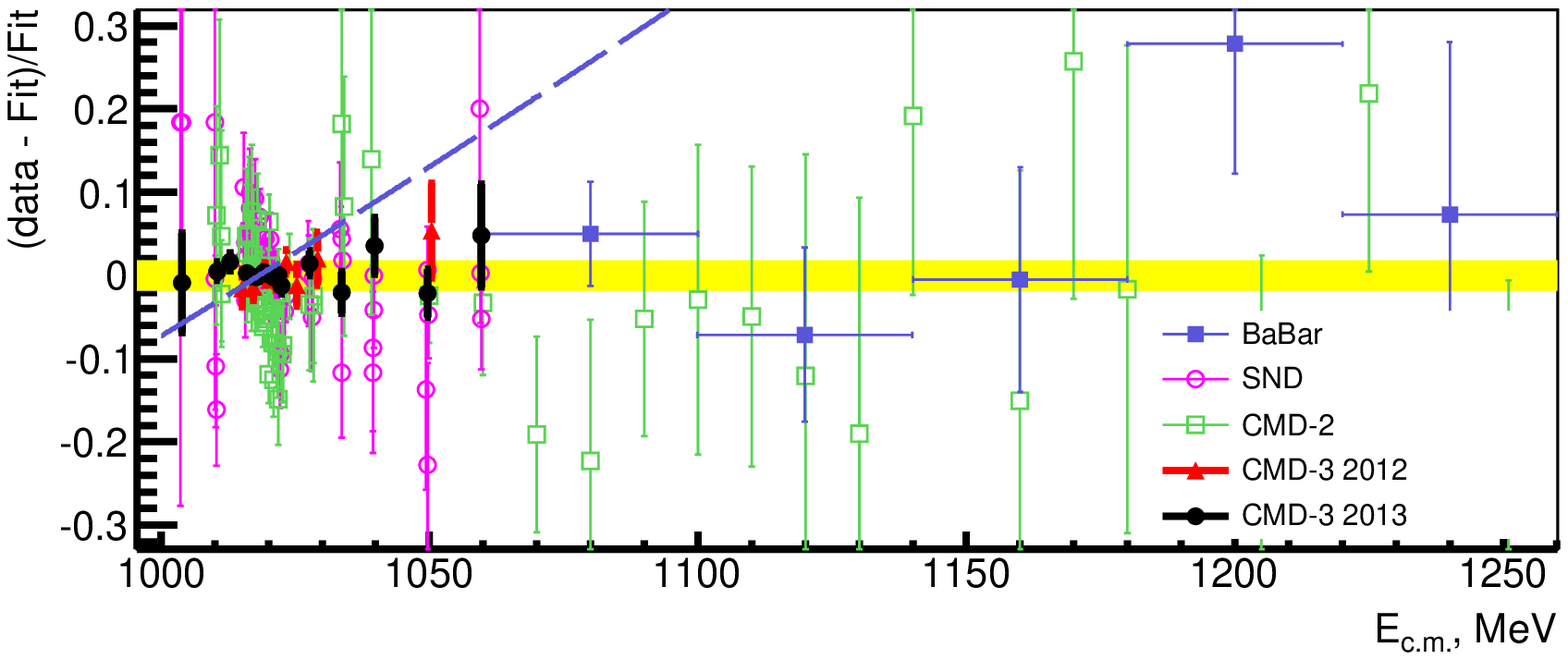}
  \caption{Relative difference between the data and fit for the 
$E_{\rm c.m.} = 1.00-1.25$ GeV range. The dashed line is the contribution of the $\phi$ meson amplitude only.}
  \label{cross_phi_rel}
\end{center}
\end{figure*}

 Figure~\ref{cross_phi_rel} shows available experimental data up to 
$E_{\rm c.m.}$ = 1250 MeV 
and demonstrates that the obtained fit parameters do not contradict 
other measurements at higher $E_{\rm c.m.}$ values. 
The dashed line shows the contribution of the $\phi$ meson only, 
when the amplitudes from the $\rho(770)$ and $\omega(782)$ are excluded 
demonstrating that the destructive  interference with these states 
dominates in the shown energy region.

\begin{table*}
\caption{The results of the approximation procedure in comparison with previous experiments}
\label{endtable_phi_parameters}
\begin{center}
\begin{tabular}{||c|c|c||}
\hline
    Parameter                                          & CMD-3  & Other measurements          \\
\hline
$m_{\phi}$, MeV                                        & 1019.457 $\pm$ 0.006 $\pm$ 0.060 $\pm$ 0.010   & 1019.461 $\pm$ 0.019 (PDG2014) \\
$\Gamma_{\phi}$, MeV                                   & 4.240    $\pm$ 0.012 $\pm$ 0.005 $\pm$ 0.010  & 4.266 $\pm$ 0.031 (PDG2014)\\
$\Gamma_{\phi \to ee} B_{\phi \to K^0_S K^0_L}$, keV   & 0.428   $\pm$ 0.001 $\pm$ 0.008 $\pm$ 0.005  & 0.4200 $\pm$ 0.0127  (BaBar)\\
$B_{\phi \to ee} B_{\phi \to K^0_S K^0_L}, 10^{-5}$    & 10.078  $\pm$ 0.025 $\pm$ 0.188 $\pm$ 0.118  & 10.06  $\pm$ 0.16 (PDG2014)\\
\hline
\hline
\end{tabular}
\end{center}
\end{table*}

\section{Conclusion}
Using the $K^0_{S}\to\pi^+\pi^-$ decay  we observe 6.1$\times$10$^5$ events of 
the process  $e^+ e^- \to K^0_{S}K^0_{L}$  in the 1004--1060 MeV c.m. energy 
range, and measure the  cross section with a 1.8\% systematic uncertainty. 
The following values of the $\phi$ meson parameters have been obtained:
$$m_{\phi} = 1019.457 \pm 0.061~ \rm  MeV/c^2  $$
$$\Gamma_{\phi} = 4.240 \pm 0.017~\rm  MeV$$
$$\Gamma_{\phi \to ee} B_{\phi \to K^0_S K^0_L} = 0.428 \pm 0.009~\rm keV .$$
The obtained parameters are in good agreement with previous experiments.
The values of $\Gamma_{\phi}$ and $\Gamma_{\phi \to ee} B_{\phi \to K^0_S K^0_L}$ are the most precise among all existing measurements.
High precision in the cross section measurement allows the first quantitative 
estimate of the contributions from $\rho$ and $\omega$ mesons to the studied c.m. region.

\section{Acknowledgements}
We thank the VEPP-2000 personnel for excellent machine operation.
This work is supported in part by the Russian Education and
by the Russian Foundation for Basic RFBR 14-02-00580-a, RFBR 14-02-91332,
RFBR 15-02-05674-a, RFBR 16-02-00160-a and the DFG grant HA 1457/9-1.

\begin{table*}
\small
\caption{The c.m. energy $E_{\rm c.m.}$, number of selected signal events $N$,  
detection efficiency $\epsilon_{\rm MC}$, 
radiative-correction factor 1 + $\delta_{\rm rad.}$, integrated luminosity $L$, 
and Born cross section $\sigma$ of the process $e^+ e^- \to K^0_{S}K^0_{L}$. 
Statistical errors only are shown.
}
\label{endtable}
\begin{center}

\begin{tabular}[t]{||c|c|c|c|c|c|c|c||}
\hline
  & $E_{\rm c.m.}$, MeV     &    $N$ events    &   $\epsilon_{\rm MC}$       & 1 + $\delta_{\rm rad.}$  &     $ 1 + \delta_{\rm en.spr.}$  &  $L$, nb$^{-1}$ &   $\sigma$, nb \\
\hline
1 & 1004.066 $\pm$ 0.008 & 315 $\pm$ 19 & 0.321 & 0.72 & 0.994 & 195.35 $\pm$ 0.67 & 6.87 $\pm$ 0.42 \\
2 & 1010.466 $\pm$ 0.010 & 9083 $\pm$ 100 & 0.312 & 0.73 & 0.992 & 936.05 $\pm$ 1.44 & 42.16 $\pm$ 0.47 \\
3 & 1012.955 $\pm$ 0.007 & 10639 $\pm$ 108 & 0.308 & 0.72 & 0.988 & 485.35 $\pm$ 1.04 & 96.74 $\pm$ 1.00 \\
4 & 1015.068 $\pm$ 0.012 & 2347 $\pm$ 50 & 0.307 & 0.71 & 0.987 & 47.91 $\pm$ 0.33 & 219.53 $\pm$ 5.02 \\
5 & 1016.105 $\pm$ 0.010 & 15574 $\pm$ 130 & 0.304 & 0.71 & 0.978 & 192.11 $\pm$ 0.66 & 366.33 $\pm$ 3.33 \\
6 & 1017.155 $\pm$ 0.012 & 65612 $\pm$ 264 & 0.303 & 0.70 & 0.983 & 478.99 $\pm$ 1.04 & 628.15 $\pm$ 2.95 \\
7 & 1017.156 $\pm$ 0.013 & 5525 $\pm$ 77 & 0.302 & 0.70 & 0.985 & 40.76 $\pm$ 0.3 & 624.76 $\pm$ 9.89 \\
8 & 1018.046 $\pm$ 0.021 & 102233 $\pm$ 334 & 0.301 & 0.70 & 0.992 & 478.34 $\pm$ 1.04 & 996.62 $\pm$ 4.28 \\
9 & 1019.118 $\pm$ 0.016 & 98014 $\pm$ 326 & 0.3 & 0.72 & 1.028 & 328.62 $\pm$ 0.86 & 1413.65 $\pm$ 6.02 \\
10 & 1019.214 $\pm$ 0.019 & 16059 $\pm$ 132 & 0.299 & 0.72 & 1.022 & 52.75 $\pm$ 0.34 & 1433.05 $\pm$ 15.03 \\
11 & 1019.421 $\pm$ 0.028 & 11066 $\pm$ 110 & 0.299 & 0.73 & 1.024 & 36.04 $\pm$ 0.28 & 1434.84 $\pm$ 18.40 \\
12 & 1019.902 $\pm$ 0.012 & 140758 $\pm$ 386 & 0.299 & 0.75 & 1.016 & 472.34 $\pm$ 1.04 & 1341.91 $\pm$ 4.74 \\
13 & 1021.222 $\pm$ 0.021 & 47552 $\pm$ 225 & 0.299 & 0.83 & 0.994 & 228.34 $\pm$ 0.72 & 833.20 $\pm$ 4.89 \\
14 & 1021.309 $\pm$ 0.009 & 9545 $\pm$ 102 & 0.299 & 0.83 & 0.994 & 46.85 $\pm$ 0.33 & 807.54 $\pm$ 10.36 \\
15 & 1022.078 $\pm$ 0.021 & 31323 $\pm$ 183 & 0.297 & 0.88 & 0.989 & 201.61 $\pm$ 0.68 & 582.93 $\pm$ 4.03 \\
16 & 1022.744 $\pm$ 0.019 & 14517 $\pm$ 126 & 0.297 & 0.93 & 0.989 & 116.71 $\pm$ 0.52 & 443.71 $\pm$ 4.38 \\
17 & 1023.264 $\pm$ 0.025 & 6876 $\pm$ 86 & 0.297 & 0.96 & 0.992 & 62.91 $\pm$ 0.38 & 377.77 $\pm$ 5.31 \\
18 & 1025.320 $\pm$ 0.031 & 2319 $\pm$ 51 & 0.294 & 1.08 & 0.996 & 36.32 $\pm$ 0.28 & 199.26 $\pm$ 4.97 \\
19 & 1027.956 $\pm$ 0.015 & 8150 $\pm$ 94 & 0.294 & 1.21 & 0.997 & 195.83 $\pm$ 0.67 & 115.93 $\pm$ 1.70 \\
20 & 1029.090 $\pm$ 0.014 & 1911 $\pm$ 45 & 0.293 & 1.26 & 0.998 & 52.94 $\pm$ 0.35 & 96.96 $\pm$ 3.00 \\
21 & 1033.907 $\pm$ 0.011 & 3704 $\pm$ 64 & 0.292 & 1.43 & 0.999 & 175.55 $\pm$ 0.64 & 50.12 $\pm$ 1.26 \\
22 & 1040.028 $\pm$ 0.035 & 2839 $\pm$ 56 & 0.289 & 1.6 & 1 & 195.91 $\pm$ 0.68 & 31.27 $\pm$ 1.01 \\
23 & 1049.864 $\pm$ 0.011 & 4291 $\pm$ 70 & 0.284 & 1.78 & 1 & 499.59 $\pm$ 1.09 & 16.93 $\pm$ 0.50 \\
24 & 1050.862 $\pm$ 0.031 & 1310 $\pm$ 39 & 0.285 & 1.79 & 1 & 146.31 $\pm$ 0.59 & 17.47 $\pm$ 0.94 \\
25 & 1059.947 $\pm$ 0.015 & 1271 $\pm$ 38 & 0.276 & 1.91 & 1 & 198.86 $\pm$ 0.69 & 12.09 $\pm$ 0.71 \\

\hline
\end{tabular}
\normalsize
\end{center}
\end{table*}

\begin{thebibliography}{0}
\bibitem{g2}
T. Blum {\it et al.}, arXiv:1311.2198.
\bibitem{snd}
M.N. Achasov et al., Phys. Rev. D {\bf 91}, 092010 (2015).
\bibitem{cmd}
D.N. Shemyakin et al., Phys. Lett. B {\bf 756}, 153 (2016).
\bibitem{Bramon}
  A. Bramon {\it et al.}, Phys. Lett. B {\bf 486}, 406 (2000).  
\bibitem{cmdn}
R. R. Akhmetshin {\it et al.} (CMD-2 Collaboration), 
Phys. Lett. B {\bf 695}, 412 (2011).
\bibitem{sndn}
 M. N. Achasov {\it et al.} (SND Collaboration), 
Phys. Rev. D {\bf 63}, 072002 (2001).
\bibitem{babarn}
 J. P. Lees {\it et al.} (BaBar Collaboration), 
Phys. Rev.D {\bf 89}, 092002 (2014). 
\bibitem{cmd3}
 B. I. Khazin {\it et al.}, Nucl. Phys. B (Proc. Suppl.) {\bf 376}, 181 (2008).
\bibitem{vepp2000000}
Yu. M. Shatunov {\it et al.}, in Proceedings of the 7th European Particle 
Accelerator Conference, Vienna, 2000, p. 439. 
\bibitem{csi}
V. M. Aulchenko {\it et al.}, JINST {\bf 10}, P10006 (2015). 
\bibitem{GEANT4}
S. Agostinelli {\it et al.} (GEANT4 Collaboration), 
Nucl. Instr. and Meth. A {\bf 506}, 250 (2003).
 \bibitem{PJGen_sibid}
  A. B. Arbuzov {\it et al.}, Eur. Phys. J. C {\bf 46}, 689 (2006).
\bibitem{rmc}
S. Actis {\it et al.} (Working Group for Radiative Corrections and Monte Carlo
Generators at Low Energies),  Eur. Phys. J. C {\bf 66}, 585 (2010). 
\bibitem{gener}
H. Czy\.{z} {\it et al.}, arXiv:1312.0454.
\bibitem{compton}
E. V. Abakumova {\it et al.}, Nucl. Instrum. Methods Phys. Res., 
Sect. A {\bf 744}, 35 (2014).
\bibitem{compton1}
E. V. Abakumova {\it et. al.}, JINST {\bf 10}, T09001 (2015).
 \bibitem{PDG}
K. A. Olive {\it et al.} (Particle Data Group), 
Chin. Phys. C {\bf 38}, 090001 (2014). 
\bibitem{lum}
G. V. Fedotovich {\it et al.}, Physics of Atomic Nuclei {\bf 78}, 591 (2015).
\bibitem{radcorFadin}
E. A. Kuraev and V. S. Fadin, Sov. J. Nucl. Phys. {\bf41}, 466 (1985).
\bibitem{Kuhn}
C. Bruch, A. Khodjamirian and J. K\"{u}hn, Eur. Phys. J. C {\bf 39}, 41 (2005).


\end{thebibliography}
\end{document}